\documentclass{vldb}

\usepackage{balance}  %
\usepackage{booktabs}
\usepackage{url}
\usepackage{dblfloatfix}
\usepackage{float}

\graphicspath{{./figures/}}
\usepackage{graphicx}
\usepackage[font=bf,labelfont=bf]{caption}
\usepackage{subcaption}
\usepackage{tikz}
\usetikzlibrary{positioning,shadows,shapes,arrows}
\usetikzlibrary{arrows.meta,chains,matrix,decorations.pathreplacing}
\usepackage{pgfplots}
\usetikzlibrary{arrows.meta,
                calc, chains,
                positioning,
                shapes.multipart}
\usepackage{listings}

\usepackage{caption}

\newlength\myindent
\usetikzlibrary{matrix,positioning,arrows.meta,arrows}
\usepackage{pdfpages}

\pgfplotsset{compat=1.16}
\usetikzlibrary{decorations.pathreplacing,calligraphy}

\usepackage{algorithm}
\usepackage{algorithmic}

\usepackage{xspace}
\newcommand*{\eg}{e.g.,\@\xspace}
\newcommand*{\ie}{i.e.,\@\xspace}
\newcommand*{\cf}{cf.\@\xspace}

\usepackage{xcolor}

\newcommand{\sparagraph}[1]{\vspace{1mm}\noindent {\bf #1}}

\vldbTitle{Towards Practical Learned Indexing}
\vldbAuthors{Mihail Stoian, Andreas Kipf, Ryan Marcus, Tim Kraska}

\begin{document}

\title{Towards Practical Learned Indexing (Extended Abstracts)}

\numberofauthors{4} %

\author{
\alignauthor
Mihail Stoian\\
\affaddr{TUM}\\
\email{mihail.stoian@tum.de}

\alignauthor
Andreas Kipf\\
\affaddr{MIT CSAIL}\\
\email{kipf@mit.edu}

\alignauthor
Ryan Marcus\\
\affaddr{MIT CSAIL, Intel Labs}\\
\email{ryanmarcus@mit.edu}
\and
\alignauthor
Tim Kraska\\
\affaddr{MIT CSAIL}\\
\email{kraska@mit.edu}
}

\maketitle

\begin{abstract}

Latest research proposes to replace existing index structures with learned models.\ However, current learned indexes tend to have many hyperparameters, often do not provide any error guarantees, and are expensive to build. We introduce Practical Learned Index (PLEX).\
PLEX only has a single hyperparameter $\epsilon$ (maximum prediction error) and offers a better trade-off between build and lookup time than state-of-the-art approaches.\ Similar to RadixSpline, PLEX consists of a spline and a (multi-level) radix layer.\ It first builds a spline satisfying the given $\epsilon$ and then performs an ad-hoc analysis of the distribution of spline points to quickly tune the radix layer.

\end{abstract}
\section{Introduction} \label{sec:introduction}

We introduce Practical Learned Index (PLEX).
Compared to existing learned indexes, PLEX only has a single hyperparameter $\epsilon$ (maximum prediction error) and is hence easy to use.
PLEX builds upon RadixSpline~\cite{radixspline} and CHT~\cite{cht}, and uses a spline layer to ensure an $\epsilon$ error bound and a radix layer that is inspired by CHT.

\sparagraph{RadixSpline.}\ RadixSpline (RS)~\cite{radixspline} consists of a linear spline model that approximates the CDF of the data within an error bound and a radix table that indexes the computed spline points.\ The main feature of RS is that it can be built with a constant amount of work per new element, which allows a single-pass build phase.\ However, when the radix table is not able to properly index the spline keys (\ie in the presence of outliers), RS can have poor  performance.

PLEX addresses two problems of RS:\ (i) the radix layer is hard to parametrize and (ii) the radix layer can be affected by outliers.\
Despite the fact that PLEX needs to perform an additional pass over the spline points array, PLEX manages to retain the high build performance of RS.

\sparagraph{Hist-Tree.}\
Hist-Tree (HT)~\cite{cht} approximates the data distributed as a histogram, instead of using functions, with a fast radix tree-based traversal method to find the right histogram bucket.\ As such, HT has similarities with a traditional radix tree, as its indexing approach also uses the binary representation of the keys, and with a learned index as it tries to approximate the data distribution rather using the traditional key-comparison present in B-tree traversals.\ HT recursively splits the data into bins until a certain threshold for the number of elements is reached.\ Apart from allowing updates, Hist-Tree also comes in a \emph{compact} version (CHT), optimized for read-only workloads.\ CHT is essentially a lookup table which stores the nodes of the initial HT and their prefix sums.\ CHT can either be built from a HT, by running a depth-first pre-order tree traversal, or, as we have implemented it in this work, directly from the data itself.\ The latter is more advantageous for our case, as keys can be processed in chunks.\ While CHT did not have an auto-tuner, we now introduce one as part of this work (auto-tuning CHT is part of our auto-tuning process).

\sparagraph{Other Related Work.}\
The first learned index, which paved the way for the development of new alternatives of index structures, is the recursive model index (RMI)~\cite{learnedindexes}.\ RMIs use learned models, arranged in a hierarchical structure, which are trained via supervised learning techniques on the cumulative distribution function (CDF) of the underlying data.\ Since then, a large suite of learned index structures have been proposed.\ Few of them support writes:\ PGM-index, a multi-level structure, where each level represents an error-bounded piecewise linear regression~\cite{pgm} or ALEX~\cite{alex}, which combines insights from RMI with proven storage and indexing techniques.\ With regard to multi-dimensional data, there exist several other recent approaches, including Flood~\cite{flood} and Tsunami~\cite{tsunami}.

\section{PLEX}
\label{sec:triespline}

PLEX is a combination of CHT~\cite{cht} and RS~\cite{radixspline}.
We build a spline on the underlying data, and then index the spline points in a CHT.
The resulting index structure features fast build times, error-bounded lookups, and is easy to use as it has only one hyperparameter, the maximum error $\epsilon$.

RS employs a radix table for indexing its spline points. However, when the keys of the dataset cannot be easily indexed by a radix table, \ie the longest common prefixes of their binary representation are large, RS can have a decrease in performance. PLEX addresses this issue by replacing the radix table with a radix tree, represented by CHT. Like RS, PLEX is built ``bottom-up'', by first constructing an error-bounded spline, which is then indexed in CHT.

\sparagraph{Build Spline.}\ The core part of PLEX is the linear spline model $S$, which approximates the position of each key $k$ within $\epsilon$ positions, $\epsilon > 0$.\ Formally, if $p^{*}$ is the position of $k$ in the CDF and $\tilde{p}$ is the approximated position computed by the spline, then $|\tilde{p} - p^*| \leq \epsilon$.\ In other words, the spline model \emph{always} predicts the correct location of the data within a maximum error of $\epsilon$.

The error-bounded spline model is defined as a set of connected linear spline points, which are picked \emph{from} the set of CDF points.\ An optimal spline, \ie the one with the fewest number of spline points, can be computed via dynamic programming in $O(N^2)$, where $N$ is the CDF size.\ Since this approach does not scale for large datasets, we use instead a greedy algorithm, which does not guarantee optimality anymore, but can be implemented in $O(N)$ \cite{Neumann2008c}.

A lookup consists of first determining in which spline segment $\sigma$ the key $k$ is located, \ie between which spline nodes the key lies in, and then performing a linear interpolation in the respective segment.\ For more details on the error-bounded spline algorithm, we refer the reader to \cite{Neumann2008c}.

\sparagraph{Build CHT.}\ Once the spline points have been selected, they can be indexed in CHT, using our new implementation, which iteratively builds each level of the tree by analyzing chunks of keys.\ This is different from the proposed bulk-loading in \cite{cht}, as our method \emph{directly} builds CHT instead of first building a sparse tree.

CHT has two hyperparameters: the number of radix bits $r$ of each tree node, \ie the fanout of the tree equals $2^r$, and the error $\delta$ within the index approximates the positions of the keys.\ Formally, if $q^{*}$ is the position of a spline point and $\tilde{q}$ is its estimated position computed by CHT, then $q^{*} \in \{\tilde{q}, \dotsc, \tilde{q} + \delta - 1\}$.\ This is a generalization of the radix table of RS, since setting $\delta = \infty$ leads to a CHT with a single node (a radix table).\
Notably, a radix table does not have a global bounded error (it must instead use the position of the \textit{next} prefix to obtain an upper bound on $\tilde{q}$).\
This requires us to develop separate cost models for each of them (\cf Section~\ref{sec:tuning}).

\sparagraph{Lookup.}\ A lookup for key $k$ starts by searching the position of the spline segment $\sigma$.\ This routine is done in two steps: First, a lookup in CHT returns an approximated position $\tilde{q}$.\ Next, a binary search is performed in the range $\{\tilde{q}, \dotsc, \tilde{q} + \delta - 1\}$ to find the exact position of $\sigma$.\ Subsequently, we perform a linear interpolation between the two spline points of $\sigma$ to obtain an estimated CDF position $\tilde{p}$ of the key.\ Finally, we perform a binary search within the error bounds $\tilde{p} \pm \epsilon$ to find the first occurrence of $k$.

\section{Auto Tuning}
\label{sec:tuning}

One main drawback of (learned) index structures is the choice of hyperparameters, as they have to be manually tuned in order to find the best lookup time under certain constraints (\eg space). %

We introduce cost models that approximate the lookup times and accurately compute the space consumption for both radix table and CHT without building the actual data structures.

\sparagraph{Radix Table.}\ A radix table with parameter $r$ splits the input data into $2^{r}$ buckets based on the first $r$ most significant bits (prefix) of the keys.\ In RS, the input data for the radix table are the keys of the spline points.\ Therefore, a lookup for key $k$ consists of first finding out in which radix bucket $b_k$ the key is located, and then performing a local search on the spline nodes within the bucket to find the exact spline segment.\ If the local search is implemented as a binary search, then the number of steps equals $\lceil{log_2(|b_k|)}\rceil$, where $|b_k|$ is the number of spline points within bucket $b_k$.

Assume that only positive lookups are performed, \ie the lookup key lies within the stored data $D$. Then the average lookup time $\lambda_r$ can be estimated as

\begin{align}
\lambda_r = \frac{1}{|D|}\displaystyle\sum_{k \in D} \lceil{log_2(|b_k|)}\rceil.\label{eq:diff}
\end{align}

This cost model has the advantage that it can be computed for all $r$ while building the spline model, without storing the actual radix tables.\ It also allows us to \textit{detect} the outlier problem of RS:\ Consider, for example, $r = 1$.\ The optimal $\lambda_1$ is achieved when the radix table splits the set of spline points in two equal-sized buckets, \ie $\lambda_1 = \lceil{log_2(\frac{|S|}{2})}\rceil = \lceil{log_2(|S|)}\rceil - 1$.\ Note that we increased the number of bits and the cost decreased by $\lambda_0 - \lambda_1 = 1$ unit ($r=0$ corresponds to a simple binary search, \ie $\lambda_0=\lceil{log_2(|S|)}\rceil$).\ This is not always the case: when all spline keys have the same value for the most-significant bit, then the radix table is not able to split the initial bucket and we have $\lambda_1 = \lambda_0$.\ In general, $\lambda_{r} - \lambda_{r+1} \in [0,1]$ tells us whether it is worth increasing the number of radix bits by one.\ Since we must update the model for each radix table of parameter $r$ as each spline node is computed, the time complexity is $O(r^{+}|S|)$, where $r^{+}$ is the maximum $r$ allowed.%

Finally, the memory consumption of a radix table with parameter $r$ is $O(2^r)$, as we only need to store a value per bucket, namely the position of the first spline point with that prefix.

\sparagraph{CHT.}\ The same reasoning can be applied for CHT.\ The aim is to consider a large set of CHTs with different configurations $(r, \delta)$ and estimate their average lookup time and memory consumption.

A lookup for key $k$ consists of first finding out in which node $v_k$ of CHT the key is located, \ie the node which contains the bin corresponding to $k$ with size $\leq \delta$, and then performing a local search on the spline nodes within that bin.\ If the local search is implemented as a binary search, then the number of search steps equals $depth(v_k) + \lceil{log_2(\delta)}\rceil$.\ As the term $\lceil{log_2(\delta)}\rceil$ is the same for all lookup keys, we only have to compute the average depth of the leaf nodes.

In the cost model for radix table (Eq.~\ref{eq:diff}), the cost of each bucket is weighted by the number of data keys falling into that bucket.\ This is more difficult for CHT, as we would have to examine the original data multiple times, namely, as many times as the number of levels, to collect such statistics. Instead, we only calculate the average tree depth given lookups from the set of spline keys.\ Note that this is a simplification that does not guarantee that we can still accurately model the average lookup time for the data itself.\ This was not a problem for the radix table, where there is only one radix level to consider, and this one level can be covered when building the spline.\ Thus, the average lookup time can be estimated as \begin{align}
\lambda_{(r,\delta)} = \lceil{log_2(\delta)}\rceil + \frac{1}{|S|}\displaystyle\sum_{k \in S} depth(v_k).\label{eq:lambda}
\end{align}

\begin{figure}
\begin{tikzpicture}
\tikzstyle{place}=[]
\begin{scope}[cell/.style={rectangle,draw=black},
space/.style={minimum height=1.5em,matrix of nodes,row sep=-\pgflinewidth,column sep=-\pgflinewidth,column 1/.style={font=\ttfamily}}]

\node [place] (s1c) {} node[above=20mm]{\textbf{Spline keys}};
\matrix (first) [space, column 1/.style={font=\ttfamily},column 2/.style={nodes={cell,minimum width=2em}}]
{
0 & 0000 \\
1 & 0101 \\
2 & 0110 \\
3 & 0111\\
4 & 1000\\
5 & 1010\\
6 & 1011\\
7 & 1111\\
};
\end{scope}

    \begin{scope}[xshift=2.5cm,yshift=-0.5cm,node distance=0pt,
    start chain = A going right,
    arrow/.style = {draw=#1,-{Stealth[]},
    shorten >=1mm, shorten <=1mm}, %
    arrow/.default = black,
    X/.style = {rectangle, draw,%
                minimum width=2ex, minimum height=3ex,
                outer sep=0pt, on chain},
    B/.style = {decorate,
                decoration={brace, amplitude=5pt,
                pre=moveto,pre length=1pt,post=moveto,post length=1pt,
                raise=1mm,
                            #1}, %
                thick},
   B/.default = mirror, %
    C/.style = {decorate,
                decoration={brace, mirror,amplitude=2pt,
                pre=moveto,pre length=1pt,post=moveto,post length=1pt,
                raise=1mm,
                            #1}, %
                thick},]

    \begin{axis}[
    ybar,
    bar width=6pt,
    height=3cm,
    width=4cm,
    xlabel={\footnotesize position},
    ylabel={\footnotesize lcp-length},
    xtick={1,2,3,4,5,6,7},
    ytick={0,1,2,3},
    title={\textbf{LCP}-Histogram}]
    \addplot coordinates {
        (1,1) (2,2) (3,3) (4,0) (5,2) (6,3) (7,1)};
        \path (-0.75, 3) coordinate (origin);
        \path (7.5, 0) coordinate (tmp0);
        \path (7.5, 1) coordinate (tmp1);
        
        \path (9.25, 0) coordinate (tmp2);
        \path (9.25, 2) coordinate (tmp3);
    \end{axis}

\draw[B=,red,transform canvas={xshift = 0.85cm}] (first-3-2.west) -- coordinate[right of=5mm] (a) (first-4-2.west);

\draw[C=,purple,transform canvas={}] (tmp0) -- coordinate[] (b) (tmp1);
\draw[C=,teal,transform canvas={}] (tmp2) -- coordinate[] (c) (tmp3);

\draw[arrow,transform canvas={xshift = 0.75cm + 5mm}] (a) to [out=30, in=90] (origin);
\end{scope}

\begin{scope}[xshift=0,yshift=-3.0cm,
   level distance = 1.1cm,start chain = A going right,
    arrow/.style = {draw=#1,-{Stealth[]},
    shorten >=1mm, shorten <=1mm}, %
    arrow/.default = black,
    X/.style = {rectangle, draw,%
                minimum width=2ex, minimum height=3ex,
                outer sep=0pt, on chain},
    B/.style = {decorate,
                decoration={brace, amplitude=5pt,
                pre=moveto,pre length=1pt,post=moveto,post length=1pt,
                raise=1mm,
                            #1}, %
                thick},
   B/.default = mirror, %
    C/.style = {decorate,
                decoration={brace, mirror,amplitude=2pt,
                pre=moveto,pre length=1pt,post=moveto,post length=1pt,
                raise=1mm,
                            #1}, %
                thick},
      child/.style = {
        edge from parent path = {(\tikzparentnode.west)
            ++(0.25cm+#1*0.65cm, -0.25cm) -- (\tikzchildnode)},
    },]

\node [place] {} node[above=2.5mm] (s1c) {\textbf{CHT }($r = 1, \delta = 2$)};
\tikzstyle{bplus}=[rectangle split,rectangle split horizontal, rectangle split parts = 2, rectangle split empty part width=0.00mm, rectangle split every empty part={\hspace{0.5cm}}, draw]
\tikzstyle{every node}=[bplus]
\tikzstyle{level 1}=[sibling distance=20mm]
\tikzstyle{level 2}=[sibling distance=20mm]
\node (test1) {[0,3]\nodepart{two}[4,7]} [->]
child {
    node {[0,0]\nodepart{two}[1,3]}
    child[child=1.5] {node (A) {[1,1]\nodepart{two}[2,3]}}
}
child {node (test3) {[4,6]\nodepart{two}[7,7]}
    child[child=+0.25] {node (D){[4,4]\nodepart{two}[5,6]}}
}
;
\draw[arrow,transform canvas={xshift = 2mm},purple] (b) to [out=-30, in=0] (test1);
\draw[arrow,transform canvas={xshift = 2mm},teal] (c) to [out=-0, in=0] (test3);
\end{scope}

\begin{scope}[xshift=4cm,yshift=-5.5cm,
   level distance = 1.1cm,start chain = A going right,
    arrow/.style = {draw=#1,-{Stealth[]},
    shorten >=1mm, shorten <=1mm}, %
    arrow/.default = black,
    X/.style = {rectangle, draw,%
                minimum width=2ex, minimum height=3ex,
                outer sep=0pt, on chain},
    B/.style = {decorate,
                decoration={brace, amplitude=5pt,
                pre=moveto,pre length=1pt,post=moveto,post length=1pt,
                raise=1mm,
                            #1}, %
                thick},
   B/.default = mirror, %
    C/.style = {decorate,
                decoration={brace, mirror,amplitude=2pt,
                pre=moveto,pre length=1pt,post=moveto,post length=1pt,
                raise=1mm,
                            #1}, %
                thick},]

\node [place] {} node[above=2.5mm] (s2c) {\textbf{CHT }($r = 2, \delta = 2$)};
\tikzstyle{bplus}=[rectangle split,rectangle split horizontal, rectangle split parts = 4, rectangle split empty part width=0.00mm, rectangle split every empty part={\hspace{0.5cm}}, draw]
\tikzstyle{every node}=[bplus]
\tikzstyle{level 1}=[sibling distance=32.5mm]
\tikzstyle{level 2}=[sibling distance=13mm]
\node (test2) {[0,0]\nodepart{two}[1,3]\nodepart{three}[4,6]\nodepart{four}[7,7]} [->]
child {
    node {$\emptyset$\nodepart{two}[1,1]\nodepart{three}[2,2]\nodepart{four}[3,3]}}
child {node {[4,4]\nodepart{two}$\emptyset$\nodepart{three}[5,5]\nodepart{four}[6,6]}};

\draw[arrow,transform canvas={xshift = 2mm},teal] (c) to [out=0, in=45] (test2);
\end{scope}
\end{tikzpicture}
\caption{The lcp-histogram and depiction of strategy for computing the average tree depths of a large set of trees with different configurations $(r, \delta)$. Lcp-length $p=1$ is used by the CHT with $r=1$, but not also by the one with $r=2$, while lcp-length $p=2$ is used by both CHTs.}
\label{fig:strategy}
\end{figure}
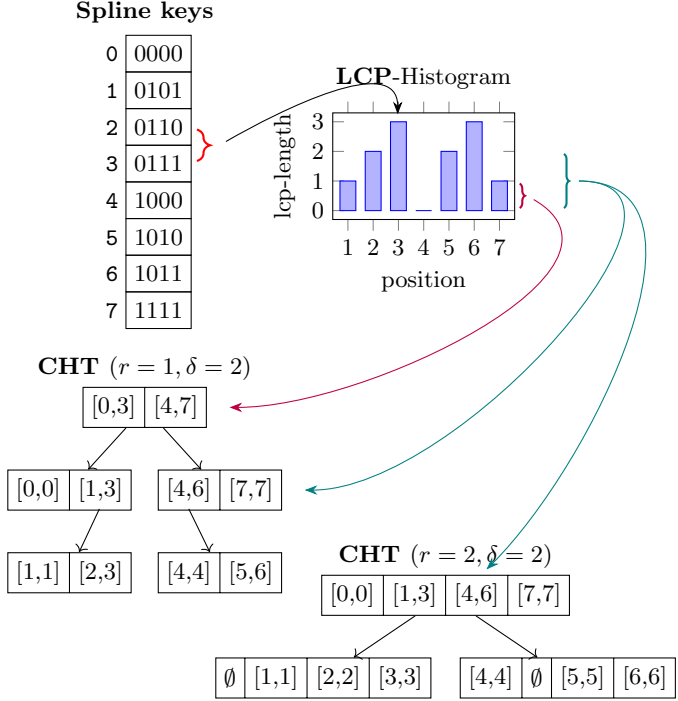

To compute the average tree depths for different pairs $(r,\delta)$, we make use of the longest common prefix (lcp) of any two adjacent spline keys.\ To this end, we build the \emph{lcp-histogram} whose values are defined as $lcp_i := lcp(keys_i,\\keys_{i-1}), \forall i \in [|S| - 1]$ (\cf Fig.~\ref{fig:strategy}).

The idea is to traverse each lcp-length $p$ of the histogram and filter out the positions whose values are smaller than $p$.\ The \emph{surviving} positions form contiguous sequences representing exactly the intervals covered by bins in particular CHTs.\ If the length of a contiguous sequence exceeds the value of $\delta$ for a particular CHT, \ie the corresponding bin of that interval is not a final one, then that bin \emph{increases} the search path by one for each key $k$ passing through it.

To illustrate this method, consider again Fig.~\ref{fig:strategy}:\ At lcp-length $p = 1$, the surviving positions, after we filter out the positions whose values are less than 1, form the contiguous sequences $1,2,3$ and $5,6,7$.\ Since $lcp$ also considers the previous key, the sequences actually define the intervals $[0,3]$ and $[4,7]$ from the set of spline keys, which can be seen in the first node of CHT$(r=1,\delta=2)$.\ Filtering out further in the same way at $p = 2$, we again get two contiguous sequences, $2,3$ and $5,6$, representing the intervals $[1,3]$ and $[4,6]$, respectively.\ These define the only two bins in both CHT$(r=1,\delta=2)$ and CHT$(r=2,\delta=2)$ that have outgoing pointers, as the length of both intervals exceeds $\delta = 2$.

In the same figure we see that lcp-length $p = 1$ is \emph{used} by the CHT with $r=1$, \ie the intervals are present in the bins of that particular CHT, but not also by the one with $r=2$, while $p=2$ is \emph{used} by both.\ Formally, a lcp-length $p$ is \emph{used} by a particular CHT$(r, \delta)$ if $p \equiv_{r} 0$.

\begin{algorithm}[t!]
\caption{Computation of average lookup times for multiple pairs $(r, \delta)$.}
\label{algo:histogram}
\begin{algorithmic}[1]
\REQUIRE upper-bounds $r^{+}, \delta^{+} \in \mathbb{N}$ on $r$ and $\delta$, respectively
\ENSURE $\lambda_{(r,\delta)}, \forall 1 \leq r \leq r^{+}, 1 \leq \delta \leq \delta^{+}$
\STATE $intervals \leftarrow$ $\{[1,|S| - 1]\}$
\FORALL{lcp-length $p$}
    \FORALL{$I$ in $intervals$}
        \FORALL{maximal subinterval $J \subseteq I$,\\s.t. $lcp_j\geq p, \forall j \in J$}
            \FORALL{$r$ in $1,\dotsc,r^{+}$ s.t. $p \equiv_{r} 0$}
                \STATE $depth_r(min(|J| - 1, \delta^{+})) \mathrel{+}= |J|$
            \ENDFOR
        \ENDFOR
    \ENDFOR
    \STATE $intervals \leftarrow$ newly found subintervals%
\ENDFOR
\FORALL {$r$ in $1,\dotsc,r^{+}$}
    \FORALL {$\delta$ in $\delta^{+} - 1,\dotsc,1$}
    \STATE $depth_r(\delta) \mathrel{+}= depth_r(\delta + 1)$
    \ENDFOR
    \FORALL {$\delta$ in $1,\dotsc,\delta^{+}$}
    \STATE $\lambda_{(r,\:\delta)} \leftarrow \lceil {log_2(\delta)}\rceil + \frac{depth_r(\delta)}{|S|}$, acc. Eq.~\ref{eq:lambda}
    \ENDFOR
\ENDFOR
\end{algorithmic}
\end{algorithm}

The method for computing $\lambda_{(r,\:\delta)}$ for different pairs $(r, \delta)$ is given in Alg.~\ref{algo:histogram}.\ The algorithm receives as input the upper-bounds $r^{+}$ and $\delta^{+}$ on $r$ and $\delta$, respectively, and outputs average lookup times for all possible pairs.\ In Line $1$, we initialize the list of intervals with the entire range of keys, \ie $[1, |S|-1]$.\ Then we iterate over all lcp-lengths $p$ and split the current intervals at positions with values~$< p$~(Line~4).\ Each new subinterval $J$ is consumed by updating $depth_r$ for all $r$ that \emph{use} the current lcp-length $p$, \ie that satisfy $p \equiv_{r} 0$. In the end, $depth_r(\delta)$ will store exactly the term $\displaystyle\sum_{k \in S} depth(v_k)$ from Eq.~\ref{eq:lambda}.%

As mentioned earlier, if the length of the interval exceeds the value of $\delta$ for a particular CHT, \ie the corresponding bin of that interval is not a final one, then that bin increases the search path by one for each key $k$ that passes through it.\ The maximum $\delta$ for which this occurs is $|J| - 1$.\ However, it would be expensive to update all $\delta \leq |J| - 1$.\ Therefore, we employ suffix sums and only update $depth_r(|J| - 1)$ in Line~6, while completing the update step for the others in Line~14.\ The algorithm finishes by computing $\lambda_{(r,\:\delta)}$ according to Eq.~\ref{eq:lambda}.\ Since we have at most $2^p$ intervals for each lcp-length $p$, the algorithm takes $O(r^{+}(|S| + d^{+}))$ time.

\begin{figure*}
\centering
\includegraphics[width=1.0\textwidth]{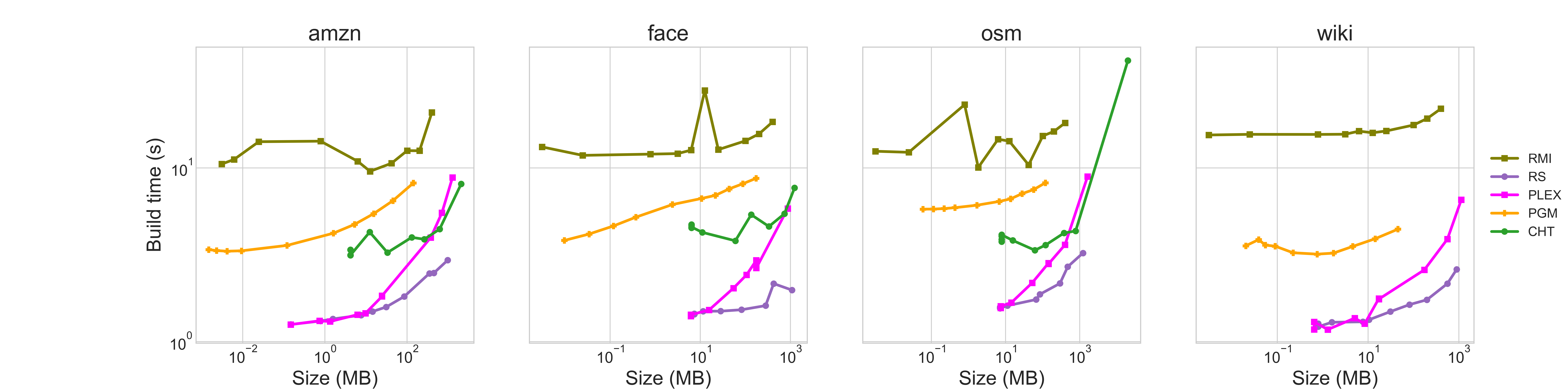}
\caption{Size versus build time on the four real-world datasets from SOSD.}
\label{fig:size_vs_build}
\end{figure*}

\begin{figure*}
\centering
\includegraphics[width=1.0\textwidth]{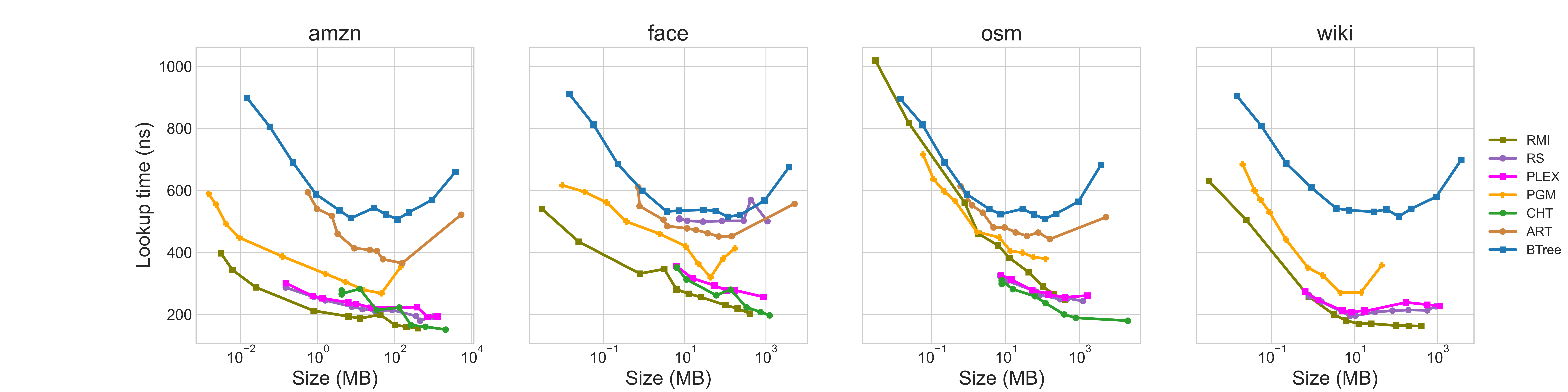}
\caption{Size versus lookup time on the four real-world datasets from SOSD.}
\label{fig:size_vs_latency}
\end{figure*}

As mentioned in~\cite{cht}, CHT can be stored as a flat array (there is no need for pointers), by representing each node as $2^r$ cells.\ Thus, the memory equals the number of nodes times $2^r$.\ The number of nodes can be computed exactly using the above strategy, since each \emph{surviving} interval represents exactly a node.\ For instance, CHT$(r=1,\delta=2$) from our example has four nodes apart from the root node, represented by the aforementioned intervals $[0,3]$, $[4,7]$, $[1,3]$, and $[4,6]$, respectively.

\sparagraph{PLEX.} The auto-tuning of PLEX relies on the optimization of both the radix table and CHT.\ After $\lambda_r$ and $\lambda_{(r,\:\delta)}$ have been computed, we can choose the best average lookup time under the constraint that the space consumption does not exceed the space requirement of the spline model, \ie in the worst case the final index structure is twice the size of the spline model.\ This way, PLEX requires only a single hyperparameter $\epsilon$.%

\section{Evaluation}\label{sec:evaluation}

\sparagraph{Setup.}\ Experiments are conducted on a m5zn.metal AWS machine with 192\,GiB of RAM and 48 vCPUs (4.5\,GHz~turbo).

\sparagraph{Datasets.} We evaluate PLEX using the SOSD benchmark \cite{sosd-neurips,sosd-vldb}.\ We use four 64-bit datasets, each of them containing 200\,M key/value pairs (3.2\,GiB):\ amzn (book popularity data), face (Facebook user IDs), osm (composite cell IDs from Open Street Map) and wiki (timestamps of Wikipedia edits).\ See~\cite{sosd-vldb} for details on these~datasets.

\sparagraph{Build Time.}\ We compare indexes regarding build time, as this highlights how expensive the learning process for each dataset is.\ We therefore consider the \textit{build time~/~size} trade-off offered by RMI, CHT, PGM, RS, and PLEX, respectively, for each dataset, as shown in Fig.~\ref{fig:size_vs_build}.\ RS achieves the lowest build times, due to its single-pass build phase. Note that the build time of PLEX already includes the auto-tuning time, unlike RS, CHT, or RMI~\cite{cdfshop}, which were tuned offline via an expensive grid search.\ Our current implementation of CHT does not support key duplicates, which is the case for the {\fontfamily{qcr}\selectfont wiki} dataset.\ PLEX does not have this problem, as the spline keys are unique.

\sparagraph{Probe Time.}\ We further compare with classical index structures, namely ART~\cite{art} and BTree~\cite{btree}.\ In the following, we consider the \emph{performance~/~size} trade-off offered by each index structure for each dataset, as shown in Fig.~\ref{fig:size_vs_latency}.\ A first observation is that PLEX successfully solves the outlier problem of RS on the {\fontfamily{qcr}\selectfont face} dataset, while preserving the performance of RS on the other datasets.\ As pointed out in \cite{sosd-vldb}, the {\fontfamily{qcr}\selectfont osm} dataset is difficult to learn, as evidenced by the behavior of RMI on this dataset. However, that is exactly where the radix approach of RS and PLEX is beneficial.\ An interesting fact to observe for both {\fontfamily{qcr}\selectfont amzn} and {\fontfamily{qcr}\selectfont face} datasets is the constant offset of $\approx{25}$ns between PLEX and RMI.\ It should also be noted that for almost all datasets, PGM, ART, and BTree eventually lose performance as the index size is increased, because the cost of navigating the index structure no longer makes up for the time required to binary search the underlying data.

\sparagraph{Auto Tuning.}\ We have empirically verified that our auto-tuning strategy succeeds in finding optimal configurations for PLEX, which have been found by grid-search on the hyperparameter triple $(\epsilon, r, \delta)$, with $\epsilon, \delta \in \{2^1,\dotsc,2^{10}\}$ and $r \in \{1, \dotsc, 10\}$.\
The auto-tuner correctly decides to fall back on the radix table as its subindex on all datasets except the {\fontfamily{qcr}\selectfont face} dataset (where it successfully detects the outlier problem and decides to use CHT).\ In some cases, it even outperforms the manually-tuned PLEX because it explored more options for $r$ and $\delta$.

\section{Conclusions}
\label{sec:conclusions}

We have introduced PLEX, an auto-tuned learned index which offers a better trade-off between build and lookup time than state-of-the-art approaches.\ PLEX is easy to use, as its only hyperparameter is the maximum prediction error.\ In future work, we plan to extend PLEX to support efficient updates by using a Fenwick tree on each tree node.%

{
\setlength{\parskip}{0.5em}
\footnotesize
\sparagraph{Acknowledgments.}\ This research is supported by Google, Intel, and Microsoft as part of DSAIL at MIT, and NSF IIS 1900933.\ This research was also sponsored by the United States Air Force Research Laboratory and the United States Air Force Artificial Intelligence Accelerator and was accomplished under Cooperative Agreement Number FA8750-19-2-1000.\ The views and conclusions contained in this document are those of the authors and should not be interpreted as representing the official policies, either expressed or implied, of the United States Air Force or the U.S. Government.\ The U.S. Government is authorized to reproduce and distribute reprints for Government purposes notwithstanding any copyright notation herein.
\par
}

\balance

\bibliographystyle{abbrv}
\bibliography{main}

\end{document}